\documentclass[12pt]{article}
\usepackage{graphicx}
\usepackage{color}
\usepackage{cite}

\def\mystrut{\vrule height 2.5ex depth 0ex width 0pt}
\def \beq{\begin{equation}}
\def \eeq{\end{equation}}
\def\eqref#1{(\ref{#1})}
\def\bea{\begin{eqnarray}}
\def\eea{\end{eqnarray}}

\def \ket#1{|{#1}\rangle}

\def\URLtilde{\lower0.2em\hbox{$\tilde{\phantom{a}}$}}
\def\mycomm#1{\hfill\break\strut\kern-3em{\color{red}\tt ====> #1
\color{black}}\hfill\break}

\newcount\timecount
\newcount\hours \newcount\minutes  \newcount\temp \newcount\pmhours
\hours = \time
\divide\hours by 60
\temp = \hours
\multiply\temp by 60
\minutes = \time
\advance\minutes by -\temp
\def\hour{\the\hours}
\def\minute{\ifnum\minutes<10 0\the\minutes
\else\the\minutes\fi}
\def\clock{
\ifnum\hours=0 12:\minute\ AM
\else\ifnum\hours<12 \hour:\minute\ AM
\else\ifnum\hours=12 12:\minute\ PM
\else\ifnum\hours>12
\pmhours=\hours
\advance\pmhours by -12
\the\pmhours:\minute\ PM
\fi
\fi
\fi
\fi
}

\def\monthname{\relax\ifcase\month 0/\or January\or February\or
March\or April\or May\or June\or July\or August\or September\or
October\or November\or December\else\number\month/\fi}

\def\bold#1{\setbox0=\hbox{$#1$}     \kern-.025em\copy0\kern-\wd0
\kern.05em\copy0\kern-\wd0
\kern-.025em\raise.0433em\box0 }

\begin{document}
\setcounter{footnote}{1}
\rightline{EFI 15-16}
\rightline{arXiv:1506.01702}
\vskip1.5cm

\centerline{\large \bf Prospects for observing the lowest-lying odd-parity
$\Sigma_c$ and $\Sigma_b$ baryons}
\bigskip

\centerline{Marek Karliner$^a$\footnote{{\tt marek@proton.tau.ac.il}}
 and Jonathan L. Rosner$^b$\footnote{{\tt rosner@hep.uchicago.edu}}}
\medskip

\centerline{$^a$ {\it School of Physics and Astronomy}}
\centerline{\it Raymond and Beverly Sackler Faculty of Exact Sciences}
\centerline{\it Tel Aviv University, Tel Aviv 69978, Israel}
\medskip

\centerline{$^b$ {\it Enrico Fermi Institute and Department of Physics}}
\centerline{\it University of Chicago, 5620 S. Ellis Avenue, Chicago, IL
60637, USA}
\bigskip
\strut

\begin{center}
ABSTRACT
\end{center}
\begin{quote}
There exist candidates for the negative-parity states $\Lambda_{c,b}(1/2^-,
3/2^-)$ consisting of an isospin-zero, spin-zero light diquark $[ud]$ with
one unit of orbital angular momentum with respect to a $c,b$ quark.  However,
there exists only one candidate for the orbital excitations of the
$\Sigma_c(1/2^+)$ and $\Sigma_c^*(3/2^+)$, and none for the orbital excitations
of $\Sigma_b(1/2^+)$ or $\Sigma_b^*(3/2^+)$.  We extend a previous
discussion of odd-parity $\Lambda_{c,b}$ states and explore some patterns of
the odd-parity $\Sigma_{c,b}$ baryons consisting of a light isospin-one
nonstrange diquark $(uu,ud,dd)$ in a state of $L=1$ with respect to the
spin-1/2 heavy quark $(c,b)$.
\end{quote}

\smallskip

\leftline{PACS codes: 12.39.Hg,12.39.Jh,14.20.Lq,14.20.Mr}
\bigskip


\section{Introduction \label{sec:intro}}

We have used simple quark-model arguments in previous work \cite{Karliner:%
2008sv,Karliner:2006ny,Karliner:2014gca} to predict the masses of some of the
lowest-lying baryons containing a single $c$ or $b$ quark.  While these
predictions generally dealt with states with no orbital excitations, the
negative-parity states $\Lambda_{c,b}(1/2^-,3/2^-)$ consisting of an 
isospin-zero, spin-zero light diquark $[ud]$ with one unit of orbital angular
momentum with respect to a $c,b$ quark were discussed briefly.  However, the
orbital excitations of the $\Sigma_{c,b}(1/2^+)$ and $\Sigma_{c,b}^*(3/2^+)$
were not treated.  Here we have labeled states with total angular momentum $J$;
the superscripts denote their parity.

In this paper we extend our previous discussion of odd-parity $\Lambda_Q$ 
states, and explore some patterns of those odd-parity $\Sigma_{c,b}$
baryons consisting of a light isospin-one nonstrange diquark $(uu,ud,dd)$
in a state of $L=1$ with respect to the spin-1/2 heavy quark $Q = (c,b)$.
This investigation is timely as a result of the demonstrated capabilities of
hadron colliders in studying heavy-hadron spectra (see, e.g., Refs.\
\cite{Aaij:2012da,Aaltonen:2013tta}.)  A comprehensive study of masses of
baryons containing heavy quarks was published several years ago
\cite{Ebert:2011kk}, but without the detail which might help identify
which of several $\Sigma_{c,b}(J^P)$ is being observed.

In Section \ref{sec:bas} we introduce basis notation for the P-wave states
of $\Lambda_Q$ and $\Sigma_Q$.  We then estimate the energy cost of a P-wave
excitation in a heavy-quark baryon by reference to existing data (Sec.\
\ref{sec:pw}), and work out spin-dependent splitting of states in the limit
where the quark $Q$ is much heavier than the diquark (Sec.\ \ref{sec:fs}).
We make some remarks on production and decay systematics in Sec.\
\ref{sec:prod} and conclude in Sec.\ \ref{sec:conc}.  An Appendix contains
details of angular momentum calculations.

\section{Basis states \label{sec:bas}}

As in Ref.\ \cite{Ebert:2011kk}, we limit our discussion to states in which a
light diquark remains in its ground state, and is orbitally excited with
respect to the heavy quark $Q$.  Internal excitation of the light
diquark seems to require more energy.  The lowest-lying negative-parity
$\Lambda$ baryons \cite{PDG}, $\Lambda(1405,1/2^-)$ and $\Lambda(1520,3/2^-)$,
may be regarded as P-wave excitations of an isospin-zero, spin-zero diquark
$[ud]$ with respect to the heavier strange quark.  The lowest P-wave $\Lambda$
state in which the $[ud]$ must be internally excited is $\Lambda(1830,5/2^-)$.
Here states are labeled by their masses in MeV.\footnote{For an extensive
discussion regarding baryons as bound states of a quark and diquark, see
Ref.\ \cite{Selem:2006nd} and references therein.}

Quantum chromodynamics (QCD) implies that the light diquark in a $\Lambda_Q$
($Q=c,b$) will be $[ud]$ in a state of zero spin and isospin, while that in a
$\Sigma_Q$ will be $(uu,ud,dd)$ in a state of unit spin and isospin.  The
diquark spin $S_d = 0$ or 1 is then coupled to the heavy quark spin $S_Q = 1/2$
and the orbital angular momentum $L=1$ to a total angular momentum $J$.

There are two convenient bases in which to evaluate the product $S_d \otimes
S_Q \otimes L$.  In the first ($L$--$S$ coupling) we couple $S_d$ and $S_Q$
to a total spin $S$ and then couple $S$ and $L$ to $J$. In the second ($j$--$j$
coupling, appropriate in the large-$m_Q$ limit \cite{DGG}), we couple
$S_d$ and $L$ to a total light-quark angular momentum $j$ and then couple $j$
and $S_Q$ to $J$.  We shall tabulate states in both bases and give the
transformation between them.

\subsection{$L$--$S$ coupling}

For $\Lambda_Q$ with an isospin-zero diquark with $S_d = 0$, the total quark
spin $S$ is necessarily 1/2.  Coupling this to the orbital angular momentum
$L=1$, one obtains states of $J^P = 1/2^-$ and $3/2^-$.  Candidates for these
states in charm and bottom sectors \cite{PDG} are summarized in Table
\ref{tab:lams}.

\begin{table}
\caption{Candidates for $\Lambda_Q$ having ground-state light diquark and heavy
quark $Q$ with relative orbital angular momentum $L=1$.
\label{tab:lams}}
\begin{center}
\begin{tabular}{c c c c c} \hline \hline
 $J^P$ & \multicolumn{2}{c}{$\Lambda_c$} & \multicolumn{2}{c}{$\Lambda_b$} \\
       & Mass (MeV) & $\Gamma$ (MeV) & Mass (MeV) & $\Gamma$ (MeV)  \\ \hline
$1/2^-$ & 2592.25$\pm$0.28 & 2.6$\pm$0.6 & 5912.1$\pm$0.4 & $<0.66$ \\
$3/2^-$ & 2628.11$\pm$0.19 & $<0.97$  &  5919.73$\pm$0.32 & $<0.63$ \\ \hline
\end{tabular}
\end{center}
\end{table}

We will anticipate a result of Sec.\ \ref{sec:fs} by noting that the
fine-structure splitting between the two $\Lambda_b$ states is less than 1/4
that between the two $\Lambda_c$ states.  This splitting should scale as
$m_c/m_b \simeq 1/3$, so the agreement is at best qualitative.

For $\Sigma_Q$ the only current candidate for a P-wave baryon is $\Sigma_c%
(2800)$ \cite{PDG}, whose spin and parity have not yet been determined.  Ref.\
\cite{Ebert:2011kk} assigns it to $J^P = 1/2^-$ or $3/2^-$ (possibly both,
overlapping in mass).  In $L$--$S$ coupling, the light $S_d=1$ diquark and
heavy $S_Q=1/2$ quark can form states with $S = 1/2$ and 3/2.  Coupling
$S=1/2$ to $L=1$ gives states with $J=1/2,3/2$, while coupling $S=3/2$ to
$L=1$ gives states with $J=1/2,3/2,5/2$.  There are thus five $\Sigma_Q$ states
with $L=1$ and odd parity (under the assumption that the diquark remains in its
ground state).  We shall introduce the notation $^{2S+1}P_J$ to describe these
states as $^2P_{1/2},~^2P_{3/2},~^4P_{1/2},~^4P_{3/2},~^4P_{5/2}$,
respectively.

The spin-dependent potential may be written \cite{Ebert:2011kk}
\beq \label{eqn:vsd}
V_{SD} = a_1{\bold L} \cdot {\bold S_d} + a_2{\bold L} \cdot {\bold S_Q}
 + b [ - {\bold S_d} \cdot {\bold S_Q}
 + ({\bold S_d} \cdot {\bold r})({\bold S_Q} \cdot {\bold r})/r^2]
 + c {\bold S_d} \cdot {\bold S_Q}~,
\eeq
where the first two terms are spin-orbit forces, the third is a tensor force,
and the last describes hyperfine splitting.  In the $L$--$S$ basis, the two
$J=1/2$ states and the two $J=3/2$ states are unmixed only if $a_1 = a_2$.
Otherwise they are eigenstates of $2 \times 2$ matrices ${\cal M}_J$ given in
the basis $[^2P_J,~^4P_J]$ by
\beq \label{eqn:m12}
{\cal M}_{1/2} = \left[ \begin{array}{c c} \frac13 a_2 - \frac43 a_1 &
\frac{\sqrt{2}}{3} (a_2-a_1) \\ \frac{\sqrt{2}}{3}(a_2-a_1) &
 - \frac53 a_1 - \frac56 a_2
\end{array} \right] +b \left[ \begin{array}{c c} 0 & 0
\\ 0 & -\frac32 \end{array} \right] + c \left[ \begin{array}{c c} -1 & 0 \\ 0 &
\frac12  \end{array} \right]~,
\eeq
\beq \label{eqn:m32}
{\cal M}_{3/2} = \left[ \begin{array}{c c} \frac23 a_1 - \frac16 a_2 &
\frac{\sqrt{5}}{3}(a_2-a_1) \\ \frac{\sqrt{5}}{3}(a_2-a_1) &
 - \frac23 a_1 - \frac13 a_2
\end{array} \right] +b \left[ \begin{array}{c c} 0 & 0
\\ 0 & \frac65 \end{array} \right] + c \left[ \begin{array}{c c} -1 & 0 \\ 0 &
\frac12 \end{array} \right]~,
\eeq
\beq
{\cal M}_{5/2} = a_1 + \frac12 a_2 - \frac{3}{10}b + \frac12 c~.
\eeq
Details of this calculation are given in Appendix A.

\subsection{$j$--$j$ coupling}

When $m_Q$ is much larger than the diquark mass, the terms $a_2$, $b$, and $c$
in \eqref{eqn:vsd} all behave as $1/m_Q$ and their expectation values are
suppressed in comparison with that of the $a_1$ term.  Thus it makes sense to
expand in a basis in which ${\bold L} \cdot {\bold S_d}$ is diagonal, treating
the other terms in $V_{SD}$ as perturbations.  Defining ${\bold j} = {\bold L}
+ {\bold S_d}$ and squaring, one finds
\beq
\langle {\bold L} \cdot {\bold S_d} \rangle = \frac12 [j(j+1) - L(L+1)
 - S_d(S_d+1)] = (-2,-1,1)~{\rm for}~j=(0,1,2)~.
\eeq
To lowest order in $1/m_Q$, the five lowest negative-parity $\Sigma_Q$ states
are displaced by $-2a_1~(J=1/2),~-a_1~(J=1/2),~-a_1~(J=3/2),~a_1~(J=3/2),~$
and $a_1~(J=5/2)$ from their spin-weighted average.  The expansion of these
states in terms of $L$--$S$ eigenstates is
\bea
\ket{J=\frac12,j=0} & = & \sqrt{\frac13} \ket{^2P_{1/2}} + \sqrt{\frac23}
 \ket{^4P_{1/2}}~, \\
\ket{J=\frac12,j=1} & = & \sqrt{\frac23} \ket{^2P_{1/2}} - \sqrt{\frac13}
 \ket{^4P_{1/2}}~, \\
\ket{J=\frac32,j=1} & = & \sqrt{\frac16} \ket{^2P_{3/2}} + \sqrt{\frac56}
 \ket{^4P_{3/2}}~, \\
\ket{J=\frac32,j=2} & = & \sqrt{\frac56} \ket{^2P_{3/2}} - \sqrt{\frac16}
 \ket{^4P_{3/2}}~, \\
\ket{J=\frac52,j=2} & = & \ket{^4P_{5/2}}~.
\eea
This allows us to evaluate the matrix elements of all the terms in the
spin-dependent potential \eqref{eqn:vsd} to lowest order in perturbation
theory by using the unperturbed eigenfunctions of the first term.  The
resultant energy shifts are
\bea
\Delta M(J=\frac12,j=0) &=& -2a_1 - b~, \\
\Delta M(J=\frac12,j=1) &=& -a_1 -\frac12 a_2 -\frac12 b -\frac12 c~, \\
\Delta M(J=\frac32,j=1) &=& -a_1 +\frac14 a_2 + b + \frac14 c~, \\
\Delta M(J=\frac32,j=2) &=& a_1 -\frac34 a_2 +\frac15 b - \frac34 c~, \\
\Delta M(J=\frac52,j=2) &=& a_1 +\frac12 a_2 - \frac{3}{10} b + \frac12 c~. \\
\eea
This expresses five mass shifts in terms of four parameters.  One linear
relation among them is the vanishing of their spin-weighted sum:
\beq \label{eqn:wtsum}
\sum_J(2J+1)\Delta M(J) = 0~.
\eeq
However, $a_2$ and $c$ always occur in the combination $a_2 + c$, so that
the five mass shifts are expressed in terms of the three free parameters $a_1$,
$a_2+c$, and $b$.  Hence the masses satisfy one additional linear relation
besides Eq.\ (\ref{eqn:wtsum}).  This is found to be
\beq \label{eqn:lin}
10 M(1/2,0) - 15 M(1/2,1) + 8 M(3/2,2) - 3 M(5/2,2) = 0~,
\eeq
where the first number refers to $J$ and the second to $j$. The mass $M(3/2,1)$
does not appear.  We shall return to this topic when discussing
numerical predictions for fine-structure splittings in Sec.\ \ref{sec:fs}.
 
\section{Energy cost of a P-wave excitation \label{sec:pw}}

One can estimate the cost of a P-wave excitation in baryons with one heavy
quark by comparing the masses of ground-state $\Lambda_Q(J^P=1/2^+)$ baryons
with those of the spin-weighted averaged masses $\bar m(\Lambda_Q)$ of the
P-wave $\Lambda_Q (J^P = 1/2^-,3/2^-)$ baryons.  The mass difference between
$\Lambda_Q(J^P=1/2^-)$ and $\Lambda_Q(J^P=3/2^-)$ is due to the ${\bold L}
\cdot {\bold S}_Q$ spin-orbit interaction, which is mathematically analogous
to the ${\bold S}_{(ud)} \cdot {\bold S}_Q$ color hyperfine interaction
responsible for $\Sigma^* - \Sigma$ splitting.  Therefore $\bar m(\Lambda_Q)
= [m(\Lambda_Q,1/2^-) + 2 m(\Lambda_Q,3/2^-)]/3$.  We shall assume that this
$\bar m(\Lambda_Q)-m(\Lambda_Q)$ splitting, denoted by
$\Delta E_{PS}$, is a function only of the reduced mass of the $Q$--diquark
system.  For that purpose we need both diquark and $Q$ masses.

The effective light isoscalar diquark mass in a $\Lambda$ is
\beq
m_{[ud]} = m_\Lambda - m_s = 577.7 {\rm MeV}~,
\eeq
where we have taken $m_\Lambda$ from Ref.\ \cite{PDG} and used $m_s = 538$ MeV
from a fit to light-quark baryon spectra \cite{Karliner:2014gca,Gasiorowicz:%
1981jz}.  (Subsequently we shall use masses from Ref.\ \cite{PDG} unless stated
otherwise.)  The corresponding light isovector diquark mass, ignoring small
isospin splittings, is
\beq \label{eqn:uudi}
m_{(uu)} = m_{(ud)} = m_{(dd)} = \frac{m_\Sigma + 2 m_{\Sigma^*}}{3} - m_s =
782.8~{\rm MeV}~.
\eeq
The spin-averaged diquark mass is $[m_{[ud]} + 3m_{(uu)}]/4 = 731.5$ MeV, close
to $2m_u=2 m_d=726$ MeV in Refs.\ \cite{Karliner:2014gca,Gasiorowicz:1981jz}.
The difference of a few MeV may be regarded as an estimate of systematic error
in our approach.

The heavy quark masses may be estimated using the difference between $\Lambda_Q$
and isoscalar diquark masses:
\beq
m_c = m_{\Lambda_c} - m_\Lambda + m_s = 1708.8~{\rm MeV}~,~~
m_b = m_{\Lambda_b} - m_\Lambda + m_s = 5041.8~{\rm MeV}~.
\eeq
These are within a couple of MeV of values estimated in Ref.\ \cite{Karliner:%
2014gca} by a slightly different method.

\begin{figure}
\begin{center}
\includegraphics[width=0.56\textwidth]{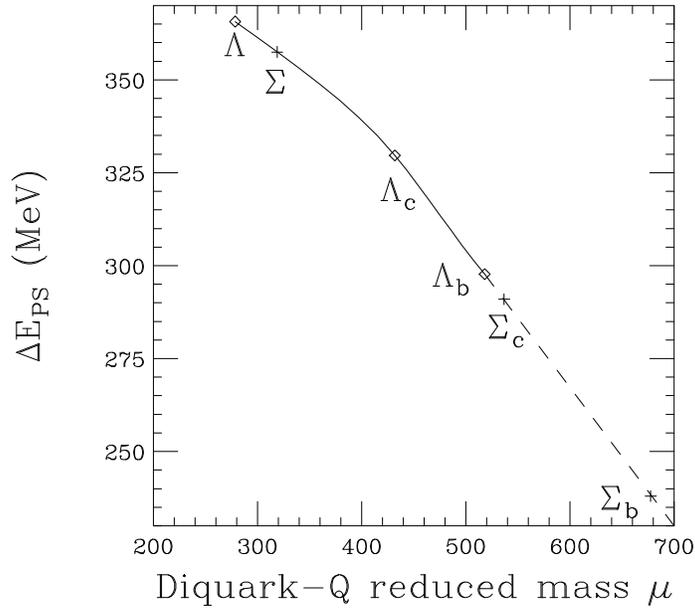}
\end{center}
\caption{Dependence of $P$--$S$ splitting parameter $\Delta E_{PS}$ on
diquark-$Q$ reduced mass $\mu$.  Diamonds denote experimental $\Lambda_Q$
points; crosses denote interpolated or extrapolated $\Sigma_Q$ points.
\label{fig:PS}}
\end{figure}

We summarize results in Table \ref{tab:PS}, and plot values of $\Delta E_{PS}$
as a function of diquark-$Q$ reduced mass $\mu$ in Fig.\ \ref{fig:PS}.  Also
shown are reduced masses for the corresponding $\Sigma_Q$ states.  We shall
use these to estimate values of $\Delta E_{PS}$ for $\Sigma_Q$ states by
linear extrapolation from those for $\Lambda_c$ and $\Lambda_b$ states.%
\footnote{We assume here that the state $\Lambda(J^P=1/2^+)$ is
$\Lambda(1405)$, which probably
has some distortion of the mass due to coupled-channel effects to 
$\bar K N$. This could influence the $\Lambda$ point in Fig.\ \ref{fig:PS}.
The $J^P$ values of $\Lambda_b(5912)$ and $\Lambda_b(5920)$ are assumed 
to be $1/2^-$ and $3/2^-$, respectively.}

\begin{table}
\caption{Input masses, values of $\Delta E_{PS}$, and diquark-$Q$ reduced
masses for baryons containing a single heavy quark $Q$.
\label{tab:PS}}
\begin{center}
\begin{tabular}{c c c r c c} \\ \hline \hline
            & \multicolumn{2}{c}{Baryon mass} & $\Delta E_{PS}$ &
 \multicolumn{2}{c}{Reduced mass $\mu$\,(MeV)} \\
   States   & $J^P=1/2^+$ & $ \bar m(\Lambda_Q)$ & (MeV) &
 $\mu(\Lambda_Q)$ & $\mu(\Sigma_Q)$ \\ \hline
 $\Lambda\phantom{_c}$  & 1115.7 & 1481.4 & 365.7 & 278.6 & 318.9 \\
$\Lambda_c$ & 2286.5 & 2616.2 & 329.7 & 431.8 & 536.8 \\
$\Lambda_b$ & 5619.5 & 5917.2 & 297.7 & 518.3 & 677.6 \\ \hline \hline
\end{tabular}
\end{center}
\end{table}

The dashed line in Fig.\ \ref{fig:PS} shows just one possible extrapolation
of the sparse information provided by the $\Lambda_Q$ states.  The predictions
for the values of $\Delta E_{PS}$ for the $\Sigma_Q$ states must thus be taken
with some care.  Using these, however, we obtain the results shown in
Table \ref{tab:pw}.

\begin{table}
\caption{Parameters leading to estimates of spin-averaged masses $\bar M$ for
P-wave excitations of a light $I=S=1$ diquark with respect to a heavy
quark $Q$.  Here $M_{0Q} \equiv [M(\Sigma_Q) + 2 M(\Sigma^*_Q)/3$.
\label{tab:pw}}
\begin{center}
\begin{tabular}{c c c c c c} \hline \hline
Heavy & $M(\Sigma_Q)$ & $M(\Sigma^*_Q)$ & $M_{0Q}$ & $\Delta E_{PS}$ &
 $\bar M$ \mystrut \\
quark $Q$ & (MeV) & (MeV) & (MeV) & (MeV) & (MeV) \\ \hline
$c$ & 2453.4 & 2518.1 & 2496.5 & 290.9 & 2787.4 \\
$b$ & 5814.3 & 5833.8 & 5827.3 & 238.8 & 6066.1 \\ \hline
\end{tabular}
\end{center}
\end{table}

\section{Fine and hyperfine structure \label{sec:fs}}

We return to the question of fine and hyperfine structure based on Eqs.\
(11--18).  First of all, the hyperfine term $c$ is expected to be negligible,
as it results from a short-distance interaction whose matrix elements
between P-wave states should be very small.  Second, we have some idea of
the magnitude of $a_2$ as a term $a_2 \langle \bold{L} \cdot \bold{S_Q}\rangle$
is responsible for the splittings between $\Lambda_c(2592,1/2^-)$ and
$\Lambda_c(2628,3/2^-)$, and between $\Lambda_b(5912,1/2^-)$ and
$\Lambda_b(5920,3/2^1)$ (see Table \ref{tab:lams}):
\beq \label{eqn:a2}
a_2=\frac23 [M(\Lambda_Q,3/2^-)-M(\Lambda_Q,1/2^-)] = \left\{ \begin{array}{c}
76.3~(\Lambda) \\ 23.9~(\Lambda_c) \\ 5.1~(\Lambda_b) \end{array} \right\}~.
\eeq
These quantities scale roughly as inverse heavy quark mass, though $a_2$ for
$b$ baryons as evaluated using the masses in Table \ref{tab:lams} is a bit
smaller than $m^b_c/m^b_b$ times $a_2$ for charmed baryons.

There are many ways to make use of Eqs.\ (11--18), but we shall mention
just two.  First, we can take the difference of two masses of states
with the same $j$ to obtain two linear combinations of $a_2$ and $b$:
\bea
M(3/2,1) - M(1/2,1) & = & \frac34 a_2 + \frac32 b~,\\
M(5/2,2) - M(3/2,2) & = & \frac54 a_2 - \frac12 b.
\eea
This allows one to extract $a_2$ and $b$, given splittings of the levels
with the same $j$.  Given the masses for $j=1$ and $j=2$, one can
use the sum rule (\ref{eqn:lin}) to solve for $M(1/2,0)$ and compare with
observation.  If the entire parameter space is mapped out for $\Sigma_c$
states, one can use the scaling relations
\beq \label{eqn:sca}
a_1(b) = a_1(c)~,~~a_2(b) = (m^b_c/m^b_b)a_2(c)~,~~b(b) = (m^b_c/m^b_b)b(c)~,
\eeq
where the superscript refers to an effective quark mass in a baryon, to predict
the splittings for $\Sigma_b$ states.

A second use of Eqs.\ (11-18) is to check the consistency of masses quoted
in Ref.\ \cite{Ebert:2011kk} with the perturbative expressions.  Two masses
are quoted for $J=1/2$ and two for $J=3/2$, as summarized in Table
\ref{tab:ebert}.

\begin{table}
\caption{Masses of lowest-lying negative-parity $\Sigma_c$ and $\Sigma_b$
baryons in the model of Ref.\ \cite{Ebert:2011kk}.
\label{tab:ebert}}
\begin{center}
\begin{tabular}{c c c} \hline \hline
$J$ & $M(\Sigma_c)$ (MeV) & $M(\Sigma_b)$ (MeV) \\ \hline
1/2 & 2713, 2799 & 6095, 6101 \\
3/2 & 2773, 2798 & 6087, 6096 \\
5/2 &    2789    &    6084    \\ \hline \hline
\end{tabular}
\end{center}
\end{table}

In Ref.\ \cite{Ebert:2011kk} it is not clear to which $j$ a state with given
$J=1/2$ or $J=3/2$ corresponds.  However, one can use the sum rule (18) to
predict $M(3/2,2)$ given either choice for $[M(1/2,0),M(1/2,1)]$.  For
$\Sigma_c$ states, choosing [2713,2799] MeV one finds $M(3/2,2) = 2903$
MeV, while [2799,2713] MeV yields $M(3/2,2) = 2634$ MeV.  Neither mass
corresponds to one of the two choices in Table \ref{tab:ebert}.  A similar
exercise for $\Sigma_b$ states predicts $M(3/2,2) = 6102$ or 6083 MeV, closer
to the predictions [6096,6087] MeV of Ref.\ \cite{Ebert:2011kk}.

We can progress further if we use an estimate of the parameter $a_1$ from the
charmed-strange meson sector.  One must view this estimate with some caution as it involves the mesons called by the Particle Data Group \cite{PDG}
$D^*_{s0}(2317)$ and $D_{s1}(2460)$, which turned out to be lighter than
expected.  (For a discussion of this point with further references, see
\cite{Cahn:2003cw}.  The masses of these states may be governed in part by
chiral dynamics lying outside the predictive power of a spin-dependent
potential \cite{chiral}). We shall assume, as we did for the $\Sigma_Q$ baryons,
that the mass eigenstates are those with definite $j$ composed of the
orbital angular momentum $L$ and the light-quark degrees of freedom (here,
the strange quark spin).  The masses of states with definite total
angular momentum $J$ and light-quark total angular momentum $j$ are
summarized in Table \ref{tab:csbar}.

\begin{table}
\caption{Masses of P-wave $c \bar s$ mesons.
\label{tab:csbar}}
\begin{center}
\begin{tabular}{c c c} \hline \hline
State & $(J,j)$ & Mass (MeV) \\ \hline
$D^*_{s0}(2317)$ & (0,1/2) & 2317.7 \\
 $D_{s1}(2460)$  & (1,1/2) & 2459.5 \\
 $D_{s1}(2536)$  & (1,3/2) & 2535.1 \\
$D^*_{s2}(2573)$ & (2,3/2) & 2571.9 \\ \hline
\end{tabular}
\end{center}
\end{table}

Repeating the steps in which one expands around eigenstates of $j$ (details are
given in the Appendix), one finds the following expressions for the meson
masses $M(J,j)$:
\bea
M(0,1/2) & = & \bar M - a_1 - a_2 - b + \frac14 c ~, \\
M(1,1/2) & = & \bar M - a_1 + \frac13 a_2 + \frac13 b - \frac{1}{12} c~,\\
M(1,3/2) & = & \bar M + \frac12 a_1 -\frac56 a_2 + \frac16 b -\frac{5}{12}c~,\\
M(2,3/2) & = & \bar M + \frac12 a_1 +\frac12 a_2 -\frac{1}{10}b +\frac14 c~.
\eea
The parameters which reproduce the masses in Table \ref{tab:csbar} are
\beq
\bar M(c\bar s) = 2513.4~{\rm MeV}~,~~
   a_1(c\bar s) = 89.4~{\rm MeV}~,~~
   a_2(c\bar s) = 40.7~{\rm MeV}~,~~
    b(c\bar s)  = 65.6~{\rm MeV}~,~~
\eeq
where we have neglected the hyperfine term for reasons mentioned earlier.

We can now relate $a_1(c\bar s)$ to the terms $a_1$ describing the coefficients
of ${\bold L} \cdot {\bold S_d}$ in $\Sigma_c$ and $\Sigma_b$, which we assume
to be equal:
\beq
a_1 = (m^m_s/m_{(uu)})a_1(c \bar s) = (483~{\rm MeV}/783~{\rm MeV})89.4~
{\rm MeV} = 55.1~{\rm MeV}~.
\eeq
Here we have used the effective mass $m^m_s = 483$ MeV of a strange quark
in a meson \cite{Karliner:2014gca}, and the mass of the $I=S=1$ light
diquark calculated in Eq.\ (\ref{eqn:uudi}).

\begin{figure}
\begin{center}
\includegraphics[width=0.7\textwidth]{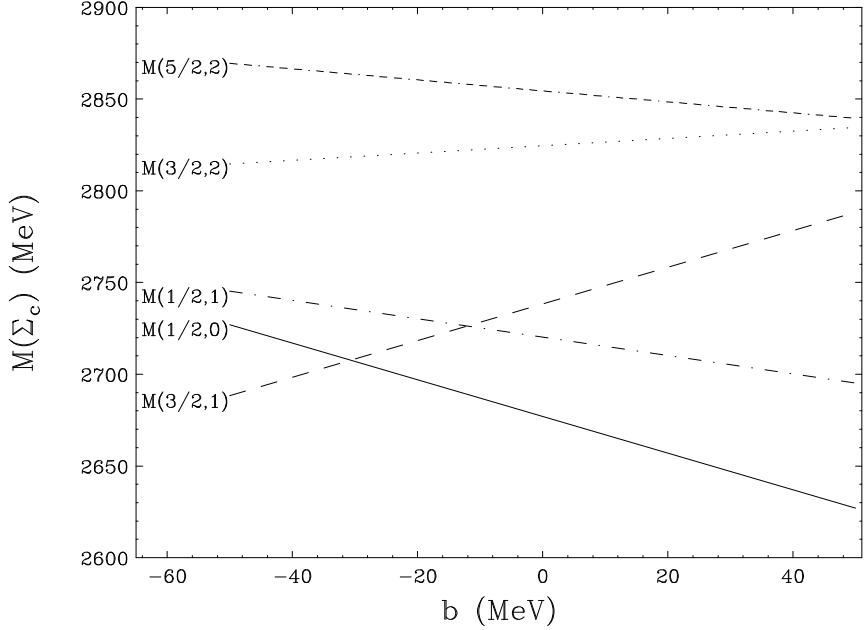}
\end{center}
\caption{Masses of P-wave $\Sigma_c$ states as functions of tensor force
parameter $b$.
\label{fig:sc}}
\end{figure}

\begin{figure}
\begin{center}
\includegraphics[width=0.7\textwidth]{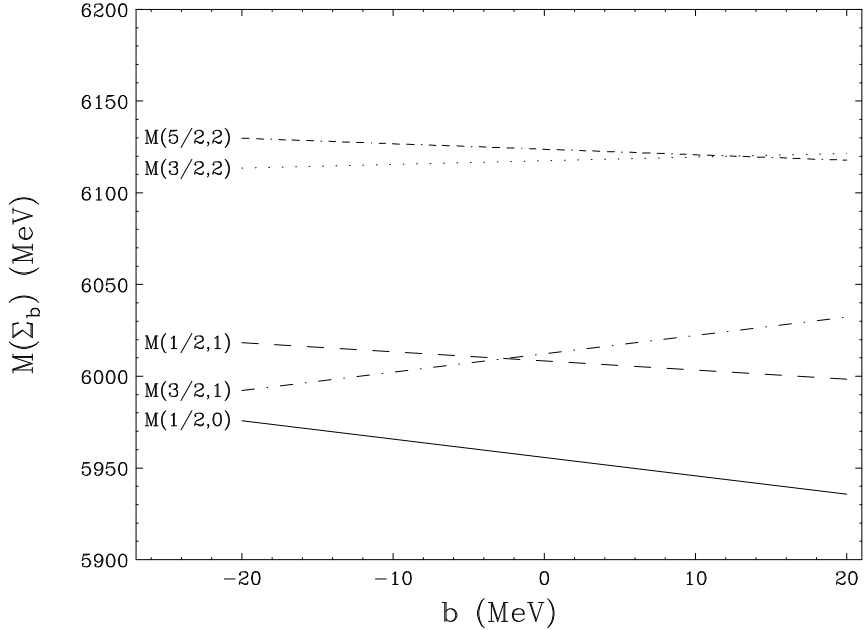}
\end{center}
\caption{Masses of P-wave $\Sigma_b$ states as functions of tensor force
parameter $b$.
\label{fig:sb}}
\end{figure}

Knowing $\bar M$, $a_1$, and $a_2$ for both $\Sigma_c$ and $\Sigma_b$
P-wave excitations, all we are missing is the tensor coefficient $b$.
We can plot predicted masses as functions of $b$ and see if one or more
solutions exist with $b$ scaling as the inverse of the heavy quark mass.
The relevant expressions, for masses $M(J,j)$ in MeV, are
\bea
\Sigma_c: M(1/2,0) & = & 2677.1 - b~,\\
          M(1/2,1) & = & 2720.3 - \frac12 b~,\\
          M(3/2,1) & = & 2738.2 + b~,\\
          M(3/2,2) & = & 2824.6 + \frac15 b~,\\
          M(5/2,2) & = & 2854.5 - \frac{3}{10}b~;
\eea
\bea
\Sigma_b: M(1/2,0) & = & 5955.8 - b~,\\
          M(1/2,1) & = & 6008.4 - \frac12 b~,\\
          M(3/2,1) & = & 6012.2 + b~,\\
          M(3/2,2) & = & 6117.4 + \frac15 b~,\\
          M(5/2,2) & = & 6123.8 - \frac{3}{10}b~.
\eea
The results are plotted as a function of the tensor parameter $b$ in
Figs.\ \ref{fig:sc} and \ref{fig:sb}.  For moderate values of $b$, there is
a clear separation between the three lowest masses $M(1/2,0)$, $M(1/2,1)$,
$M(3/2,1)$ and the two highest masses $M(3/2,2)$ and $M(5/2,2)$.  These are,
coincidentally, the states most likely to be relatively narrow and hence
easier to observe, as we shall see in the next Section.

Calculations of excited $\Sigma_c$ masses have been made in lattice QCD.
Predictions were presented in \cite{Padmanath:2013bla,Padmanath:2014bxa},
based on operators that obey an SU(3) symmetry of $u,d,c$ quarks and an SU(2)
of spin, combined into an SU(6) of which the $L=1$ baryons belong to
a 70-plet with SU(3) $\times$ SU(2) decomposition (1,2)+(8,2)+(8,4)+(10,2).
The additional $J=1/2$ and $J=3/2$ levels in \cite{Padmanath:2013bla,%
Padmanath:2014bxa} are related to spin-zero light-diquark excitations,
which have not been considered in our case.

The lattice calculations were performed at a pion mass of 400 MeV (so no
chiral extrapolation) and a single lattice spacing (not extrapolated to
continuum), yielding a calculated $\Lambda_c$ mass 149 MeV above its
experimental value.  Subtracting 149 MeV from mass values of Ref.\
\cite{Padmanath:2013bla} (Fig.\ 3), one
obtains the values, labeled by mass in MeV and total spin \cite{MPPC}:
\smallskip
\null

~~~~Octet: (2794,1/2), (2799,1/2), (2875,3/2), (2884,3/2), and (2873,5/2),
\smallskip
\null

~~~~Decuplet: (2964,1/2) and (2978,3/2),
\smallskip

\noindent
where the octet and decuplet assignments are only approximate.

Even considering that the lattice predictions are expected to be more reliable
for mass splittings than for absolute masses, this pattern is very far from
that in Fig.\ \ref{fig:sc} for any value of the parameter $b$.  It will be
interesting to compare experimental values with these two sets of predictions.

\section{Production and decay systematics \label{sec:prod}}

In the absence of further dynamical guidance, we may assume that the cross
section for production of a state with total angular momentum $J$ is
proportional to its statistical weight $(2J+1)$.  The highest-spin states are
thus most likely to be produced under such a hypothesis.  To progress further
one might have to take account of how the $I=J=1$ diquark is excited with
respect to the heavy quark $Q$.  However, there is an additional circumstance
favoring $\Sigma_Q$ states with higher spin.

For the P-wave $D$ mesons \cite{DGG}, the $D^{**}$ states with
light-quark total angular momentum $j=1/2$ are allowed to decay to $D \pi$
and/or $D^* \pi$ only via S waves, while those with $j=3/2$ can decay to
$D \pi$ and/or $D^* \pi$ only via D waves, and hence are relatively narrow.
This situation is illustrated in Table \ref{tab:cpw}, which lists the
final light-quark spins accessible when combining $j$ with the orbital
angular momentum $L$ of pseudoscalar meson emission.
\begin{table}
\caption{Values of final light-quark spin accessible in decays of P-wave
charmed mesons.
\label{tab:cpw}}
\begin{center}
\begin{tabular}{c c c c} \hline \hline
$J$ & $j$ & $L=0$ & $L=2$ \\ \hline
 0  & 1/2 & 1/2 &      3/2, 5/2      \\
 1  & 1/2 & 1/2 &      3/2, 5/2      \\
 1  & 3/2 & 3/2 & 1/2, 3/2, 5/2, 7/2 \\
 2  & 3/2 & 3/2 & 1/2, 3/2, 5/2, 7/2 \\ \hline \hline
\end{tabular}
\end{center}
\end{table}
In decays to $D \pi$ or $D^* \pi$ the final light-quark total spin is
1/2, so $j=1/2$ states decay to $D \pi$ or $D^* \pi$ only via S waves,
while $j=3/2$ states decay to $D \pi$ or $D^* \pi$ only via D waves.
The $D^{**}$ states with $j=3/2$ $D_1(2420)$ and $D_2^*(2460)$ \cite{PDG}
are thus the ones which have been firmly identified, while those with $j=1/2$,
being relatively broader, still have not been pinned down.  In the $c \bar s$
case, the $j=1/2$ states would have been broad except that they lie below
the corresponding open charm thresholds and thus must decay via isospin
violation or electromagnetically.

In the case of P-wave $\Sigma_Q$ states, a similar hierarchy holds,
illustrated in Table \ref{tab:Qpw}. 
\begin{table}
\caption{Values of final light-quark spin accessible in decays of P-wave
baryons with one heavy quark.
\label{tab:Qpw}}
\begin{center}
\begin{tabular}{c c c c} \hline \hline
$J$ & $j$ & $L=0$ & $L=2$   \\ \hline
1/2 & 0 & 0 &       2       \\
1/2 & 1 & 1 &    1, 2, 3    \\
3/2 & 1 & 1 &    1, 2, 3    \\
3/2 & 2 & 2 & 0, 1, 2, 3, 4 \\
5/2 & 2 & 2 & 0, 1, 2, 3, 4 \\ \hline \hline
\end{tabular}
\end{center}
\end{table}
The light-quark spin in a $\Lambda_Q$ is zero, so the state $(J,j) = (1/2,0)$
can decay to $\Lambda_Q \pi$ in an S wave, while $(3/2,2)$ and $(5/2,2)$ can
decay to $\Lambda_Q \pi$ in a D wave.  The light-quark spin in a $\Sigma_Q$ or
$\Sigma^*_Q$ is 1, so (1/2,1) and (3/2,1) can decay to $\Sigma^{(*)} \pi$ in
both partial waves, while (3/2,2) and (5/2,2) can decay to $\Sigma^{(*)} \pi$
only in a D wave.  Consequently, as the main decay modes of (3/2,2) and
(5/2,2) need to be in D waves, these will be the narrow states, and hence
the more easily observed of the five predicted ones.  We thus expect that the
$\Sigma_c$ state by the BaBar Collaboration at $2846 \pm 8 \pm 10$ MeV
\cite{Aubert:2008ax} is a candidate for the (5/2,2) or (3/2,2) state.
However, it is not seen by Belle \cite{Mizuk:2004yu}, who see an isotriplet
of states near 2800 MeV decaying to $\Lambda_c \pi$.

\section{Conclusions \label{sec:conc}}

We have suggested some ways to look for P-wave excitations of $\Sigma_c$
and $\Sigma_b$ baryons, concentrating on those levels in which the light
diquark in the ground state baryons with $I=J=1$ acquires one unit
of orbital angular momentum with respect to the heavy quark $Q=(c,b)$.
In this limit there are five expected P-wave $\Sigma_Q$ states:  two
with total spin $J=1/2$, two with $J=3/2$, and one with $J=5/2$.
In the heavy-quark limit one can treat $Q$ as a spectator and discuss
the total light-quark angular momentum ${\bold j} = {\bold L} + {\bold S_d}$,
where $S_d = 1$ is the spin of the diquark, so $j=0,1,2$.  We have
estimated masses as a function of one unknown parameter $b$ describing tensor
forces.  For modest values of $b$, the states with $(J,j) = (3/2,2)$ and
(5/2,2) lie highest, and these are also the ones we expect are most likely
to be detected first.  They lie somewhat above 2800 MeV for charm and 6100 MeV
for beauty.

\section*{Acknowledgements}
The work of J.L.R. was supported by the U.S. Department of Energy, Division of
High Energy Physics, Grant No.\ DE-FG02-13ER41958.  We thank M. Padmanath, Mike
Peardon, and Sheldon Stone for discussions.

\section*{Appendix:  Angular momentum state construction}

\subsection*{$L$--$S$ coupling}

In the $L$--$S$ basis, the terms proportional to $b$ and $c$ in the
spin-dependent potential (\ref{eqn:vsd}) are diagonal.  We start by showing
that the tensor force term can be written in terms of the total spin
${\bold S} = {\bold S_d} + {\bold S_Q}$:
\beq
\frac{S_{12}}{2} \equiv
\langle (3 {\bold S} \cdot {\bold r})(3 {\bold S} \cdot {\bold r})/r^2
- {\bold S}^2 \rangle = \langle 6 ({\bold S_d} \cdot {\bold r})
({\bold S_Q} \cdot {\bold r})/r^2 - 2 {\bold S_d} \cdot {\bold S_Q}\rangle~,
\eeq
which can be evaluated using an identity from Ref.\ \cite{LL}:
\beq
\langle{n_i n_j}\rangle -\frac{1}{3}\delta_{ij} = a[L_i L_j + L_j L_i - \frac23
\delta_{ij} L(L+1)]~,~~a = -1/[(2L-1)(2L+3)]~.
\eeq
In the present case for $L=1$, we have
\beq
\langle S_{12} \rangle = -\frac65 \langle [L_i L_j + L_j L_i
 - \frac43 \delta_{ij}]S_i S_j \rangle~.
\eeq
The product $L_i L_j S_i S_j$ is just $({\bold L} \cdot {\bold S})^2$, while
use of the commutation relations $[L_i,L_j] = i \epsilon_{ijk}L_k$ and
$[S_i,S_j] = i \epsilon_{ijk} S_k$ yields $L_j L_i S_i S_j = ({\bold L} \cdot
{\bold S})^2 + {\bold L} \cdot {\bold S}$.  Consequently, we have
\beq
\langle S_{12} \rangle = -\frac65 [\langle 2({\bold L} \cdot {\bold S})^2
+ {\bold L} \cdot {\bold S} \rangle - \frac43 S(S+1)]~.
\eeq
The expectation values of ${\bold L} \cdot {\bold S}$ may be evaluated, of
course, by squaring the identity ${\bold J} = {\bold L} + {\bold S}$:
\beq
\langle {\bold L} \cdot {\bold S} \rangle=\frac12 [J(J+1) - L(L+1) - S(S+1)]~.
\eeq
The values of $\langle {\bold L} \cdot {\bold S} \rangle$ and $\langle S_{12}
\rangle$ are shown in Table \ref{tab:lst}.

\begin{table}
\caption{Expectation values of ${\bold L} \cdot {\bold S}$ and tensor term
$S_{12}$ for P-wave $\Sigma_Q$ baryons in the $L$--$S$-coupling basis states.
\label{tab:lst}}
\begin{center}
\begin{tabular}{c c c} \hline \hline
State & $\langle {\bold L} \cdot {\bold S}\rangle$ & $\langle S_{12}\rangle$
 \\ \hline
$^2P_{1/2}$ & --1 & 0 \\
$^2P_{3/2}$ & $\frac12$ & 0 \\
$^4P_{1/2}$ & --$\frac52$ & --6 \\
$^4P_{3/2}$ & --1 & $\frac{24}{5}$ \\
$^4P_{5/2}$ & $\frac32$ & --$\frac65$ \\ \hline \hline
\end{tabular}
\end{center}
\end{table}

The matrix elements of ${\bold L} \cdot {\bold S_d}$ and ${\bold L} \cdot
{\bold S_Q}$ may be evaluated by
explicit construction of states with a given $J_3$ as linear combinations
of states $|S_{d3},S_{Q3},L_3 \rangle$ where $S_{d3} + S_{Q3} + L_3 = J_3$.
By angular momentum invariance, it suffices to use a single $J_3$ for each
matrix element.  An operator ${\bold L} \cdot {\bold S_i}$, where $i = d,Q$,
may be expressed as
\beq \label{eqn:ang}
{\bold L} \cdot {\bold S_{i}} = L_3 S_{i3} + \frac12 [L_+ S_{i-}+L_- S_{i+}]~,
\eeq
and the usual rules for raising and lowering third components of angular
momenta apply.  The relevant basis states are
{\small
\bea
|^2P_{1/2},J_3=\frac12 \rangle & = & \frac{\sqrt{2}}{3}\ket{1,-\frac12,0}
 -\frac13\ket{0,\frac12,0} - \frac{\sqrt{2}}{3}\ket{0,-\frac12,1}
 +\frac23\ket{-1,\frac12,1}~,\\
|^4P_{1/2},J_3=\frac12 \rangle & = & \frac{1}{\sqrt{2}}\ket{1,\frac12,-1}
 -\frac13\ket{1,-\frac12,0} - \frac{\sqrt{2}}{3}\ket{0,\frac12,0}
 +\frac13\ket{0,-\frac12,1} + \frac{1}{3\sqrt{2}}\ket{-1,\frac12,1}~,\\
|^2P_{3/2},J_3=\frac32 \rangle & = & \sqrt{\frac23}\ket{1,-\frac12,1}
 -\sqrt{\frac13}\ket{0,\frac12,1}~,\\
|^4P_{3/2},J_3=\frac32 \rangle & = & \sqrt{\frac35}\ket{1,\frac12,0}
 -\sqrt{\frac{2}{15}}\ket{1,-\frac12,1}
 -\frac{2}{\sqrt{15}}\ket{0,\frac12,1}~,\\
|^4P_{5/2},J_3=\frac52 \rangle & = & \ket{1,\frac12,1}~.
\eea
}
The matrix elements of ${\bold L} \cdot {\bold S_i}$ ($i=d,Q$) in the
basis $[^2P_J,^4P_J]$ are then found to be
\beq \label{eqn:m12dQ}
\langle{\bold L} \cdot {\bold S_d} \rangle_{J=1/2}  = \left[ \begin{array}{c c} 
-\frac43 & -\frac{\sqrt{2}}{3} \\ -\frac{\sqrt{2}}{3} & -\frac53 \end{array}
\right]~,~~ \langle{\bold L} \cdot {\bold S_Q} \rangle_{J=1/2} = \left[
\begin{array}{c c} \frac13 & \frac{\sqrt{2}}{3} \\ \frac{\sqrt{2}}{3} &
-\frac56 \end{array} \right]~,
\eeq
\beq \label{eqn:m32dQ}
\langle{\bold L} \cdot {\bold S_d} \rangle_{J=3/2}  = \left[ \begin{array}{c c} 
\frac23 & -\frac{\sqrt{5}}{3} \\ -\frac{\sqrt{5}}{3} & -\frac23 \end{array}
\right]~,~~ \langle{\bold L} \cdot {\bold S_Q} \rangle_{J=3/2} = \left[
\begin{array}{c c} -\frac16 & \frac{\sqrt{5}}{3} \\ \frac{\sqrt{5}}{3} &
-\frac13 \end{array} \right]~,
\eeq
while the matrix elements for $J=5/2$ are
\beq \label{eqn:m52dQ}
\langle{\bold L} \cdot {\bold S_d} \rangle_{J=5/2} = 1~,~~
\langle{\bold L} \cdot {\bold S_Q} \rangle_{J=5/2} = \frac12~.
\eeq
The sums of the matrix elements for ${\bold L} \cdot {\bold S_d}$ and
${\bold L} \cdot {\bold S_Q}$ reproduce those of ${\bold L} \cdot {\bold S}$
quoted in Table \ref{tab:lst}.

\subsection*{$j$--$j$ coupling}

In the limit of very large $m_Q$, the term proportional to $a_1$ in the
spin-dependent potential \eqref{eqn:vsd} dominates over the others, which may
be treated perturbatively.  Thus one can either work directly with $j$--$j$
coupling (as in Sec.\ \ref{sec:bas} or Ref.\ \cite{DGG}) or use eigenstates of
\eqref{eqn:m12dQ} and \eqref{eqn:m32dQ} to denote matrix elements of
${\bold L} \cdot {\bold S_d}$.  For $J=1/2$ the eigenvalues $\lambda$ and
corresponding eigenvectors are
\beq
\lambda = -2:~ \ket{J=1/2,j=0} = \sqrt{\frac13}\ket{^2P_{1/2}}
                               + \sqrt{\frac23}\ket{^4P_{1/2}}~,
\eeq
\beq
\lambda = -1:~ \ket{J=1/2,j=1} = \sqrt{\frac23}\ket{^2P_{1/2}}
                               - \sqrt{\frac13}\ket{^4P_{1/2}}~,
\eeq
while for $J=3/2$ they are
\beq
\lambda = -1:~ \ket{J=3/2,j=1} = \sqrt{\frac16}\ket{^2P_{3/2}}
                               + \sqrt{\frac56}\ket{^4P_{3/2}}~,
\eeq
\beq
\lambda = +1:~ \ket{J=3/2,j=2} = \sqrt{\frac56}\ket{^2P_{3/2}}
                               - \sqrt{\frac16}\ket{^4P_{3/2}}~.
\eeq
Note that these states correspond to definite values of $j$, where ${\bold j} =
{\bold L} + {\bold S_d}$.  The coefficient of $a_1$ in the spin-dependent
potential for the $J^P = \frac52^+$ state is 1, and it has $j=2$.

\subsection*{Details of calculation for P-wave $c \bar s$ mesons}

A calculation similar to that in the previous two subsections may be
performed for mesons with one heavy quark.  We chose the $c \bar s$ system
because there exist candidates for all four expected levels.  As in the
case of $\Sigma_Q$ baryons, we find it convenient to work in the $j$--$j$
basis in which the analogue of the first term in Eq.\ (\ref{eqn:vsd}), namely
$a_1(c \bar s) {\bold L} \cdot {\bold S_s}$, is diagonal.  We first calculate
the expectation values of ${\bold L} \cdot {\bold S}$ and the tensor operator
$S_{12}$ in the basis states $^{2S+1}P_J$, with $S_d$ in Eq.\ (\ref{eqn:vsd})
everywhere replaced by the spin $S_s$ of the strange quark.  The results are
shown in Table \ref{tab:lstcs}.  The expectation value of the hyperfine term
${\bold S_s} \cdot {\bold S_Q}$ is 1/4 for the $^3P$ states and --3/4 for
the $^1P_1$ state.
 
\begin{table}
\caption{Expectation values of ${\bold L} \cdot {\bold S}$ and tensor term
$S_{12}$ for P-wave $c \bar s$ mesons in the $L$--$S$-coupling basis states.
\label{tab:lstcs}}
\begin{center}
\begin{tabular}{c c c} \hline \hline
State & $\langle {\bold L} \cdot {\bold S}\rangle$ & $\langle S_{12}\rangle$
 \\ \hline
$^3P_0$ & --2 & --4 \\
$^1P_1$ & 0 & 0 \\
$^3P_1$ & --1 & 2 \\
$^3P_2$ & 1 & --$\frac25$ \\ \hline \hline
\end{tabular}
\end{center}
\end{table}

Now we construct $L$--$S$ basis states so as to evaluate the expectation
values of ${\bold L} \cdot {\bold S_i}~(i=s,Q)$.  We label the states by
$\ket{S_{s3}, S_{Q3}, L_3}$.  We do not need to exhibit the $^3P_0$ state
as it is pure $j=1/2$.  The results are
\bea
\ket{^3P_2,J_3 = 2} & = & \ket{\frac12,\frac12,1}~,\\
\ket{^3P_1,J_3 = 1} & = & \frac12\ket{-\frac12,\frac12,1}
 + \frac12 \ket{\frac12,-\frac12,1}
 -\frac{1}{\sqrt{2}} \ket{\frac12,\frac12,0}~,\\
\ket{^1P_1,J_3 = 1} & = & \frac{1}{\sqrt{2}} \ket{\frac12,-\frac12,1}
 - \frac{1}{\sqrt{2}}\ket{-\frac12,\frac12,1}~.
\eea
Using Eq.\ (\ref{eqn:ang}), in the basis $(^1P_1,^3P_1)$, the matrices
describing mixing of states are
\beq
\langle {\bold L} \cdot {\bold S_s} \rangle_{J=1} = \left[ \begin{array}{c c}
0 & 1/\sqrt{2} \\ 1/\sqrt{2} & -1/2 \end{array} \right]~,~~
\langle {\bold L} \cdot {\bold S_Q} \rangle_{J=1} = \left[ \begin{array}{c c}
0 & -1/\sqrt{2} \\ -1/\sqrt{2} & -1/2 \end{array} \right]~
\eeq
The eigenstates $[\alpha,\beta]^T$ of the first matrix satisfy
$\beta/\sqrt{2} = -\alpha$ for eigenvalue --1 and $\beta = \alpha/\sqrt{2}$
for eigenvalue 1/2.  The relation between $L$--$S$ eigenstates and $(J,j)$
eigenstates is then
\bea
\ket{J=0, j=1/2} & = & \ket{^3P_0}~,\\
\ket{J=1, j=1/2} & = & \sqrt{\frac13} \ket{^1P_1}-\sqrt{\frac23}\ket{^3P_1}~,\\
\ket{J=1, j=3/2} & = & \sqrt{\frac23} \ket{^1P_1}+\sqrt{\frac12}\ket{^3P_1}~,\\
\ket{J=2, j=3/2} & = & \ket{^3P_2}~.
\eea
Using these expressions one can calculate the expectation values of
all the operators contributing to the masses of the P-wave $c \bar s$
mesons, with the results shown in the text.

\end{document}